\documentclass[aps,prc,twocolumn,showpacs,superscriptaddress]{revtex4-1}  
\usepackage{amssymb,amsmath,mathtools}
\usepackage{graphicx}

\begin{document}
\title{Extrapolation of scattering data to the negative-energy region}

\author{L. D. Blokhintsev}
\affiliation{Skobeltsyn Institute of Nuclear Physics, Lomonosov Moscow State University, Russia}
\author{A. S. Kadyrov}
\affiliation{Department of Physics and Astronomy, Curtin University, GPO Box U1987, Perth, WA 6845, Australia}
\author{A. M. Mukhamedzhanov}
\affiliation{Cyclotron Institute, Texas A\&M University, College Station, TX 77843, USA}
\author{D. A. Savin} 
\affiliation{Skobeltsyn Institute of Nuclear Physics, Lomonosov Moscow State University, Russia}

\pacs{03.65.Nk, 03.65.Ge, 25.40.Cm, 25.55.Ci}

\begin{abstract}
Explicit analytic expressions are derived for the effective-range function 
for the case when the interaction is represented by a sum of the short-range square-well and long-range Coulomb potentials. These expressions are then transformed into forms convenient for  extrapolating to the negative-energy region and obtaining the information about bound-state properties. Alternative ways of extrapolation are discussed. Analytic properties of separate terms entering these expressions for the effective-range function and the partial-wave scattering amplitude are investigated.
\end{abstract}

\maketitle

\section{Introduction}

Asymptotic normalization coefficients (ANC) determine the asymptotics of bound-state nuclear wave functions in binary channels. ANCs are proportional to vertex constants, which determine on-shell matrix elements of virtual 
$a\leftrightarrow b+c$ processes and are related directly to the residue in energy of the elastic $b+c$ scattering amplitude at the pole corresponding to the bound state of nucleus $a$ \cite{BBD}. 
 ANCs are fundamental nuclear characteristics important both in nuclear reaction and nuclear structure physics. They are used actively in analysis of nuclear reactions within various approaches. The ANCs extracted from the analysis of one process can be used to predict features of other ones. Comparing empirical values of ANCs with theoretical ones allows one to evaluate the quality of a model. 

The ANC $C_{abc}$ for the virtual decay $a\to b+c$ determines the probability of the $\{bc\}$ configuration in nucleus $a$ at distances greater than the radius of nuclear interaction.  Thus the ANCs naturally appear in expressions for the cross sections of nuclear reactions between charged particles at low energies when, due to the Coulomb repulsion, the reactions occur at large distances. Astrophysical nuclear reactions represent the most important type of such reactions. The role of ANCs in nuclear astrophysics was first discussed in 
Refs.\cite{MukhTim,Xu}, where it was emphasized that the ANC determines the overall normalization of peripheral radiative capture reactions (see also \cite{MukhTr,Mukh2}). Thus the ANC method can be employed as an indirect technique in nuclear astrophysics. The ANCs can be used in evaluating the radiative width of a resonance, decaying to a bound state \cite{Mukh3}. An instructive example of using ANCs in nuclear structure studies is the application of the ANC to determine the radii of halo nuclei \cite{Carst}. Thus it is important to know the values of ANCs. 

In principle, values of ANCs can be deduced from the microscopic calculations of wave functions for corresponding nuclear systems. However, such calculations are quite involved  even for few-nucleon systems \cite{Quaglioni_and_Navratil}. 
On the other hand, low-energy radiative capture reactions allow us to determine ANCs experimentally.
However, ANCs, in contrast to binding energies, cannot be directly measured. Nevertheless, 
there is an indirect way to determine the ANC from experiment: the ANC $C_{abc}$ can be determined from experimental data by extrapolating, in the center-of-mass (c.m.) energy $E$,  the partial-wave amplitude of elastic $b+c$ scattering, obtained by the phase-shift analysis, to the pole corresponding to the bound state $a$ and lying at $E<0$. The most natural procedure for such extrapolation is the analytic approximation of the experimental values of the  effective-range function 
(ERF) with the subsequent continuation to the pole position. The ERF method has been successfully employed to determine the ANCs for bound (as well as resonant) nuclear states in a number of works (see, e.g. \cite{BKSSK,SpCaBa,IrOr} and references therein). 

The ERF is expressed in terms of scattering phase shifts. In case of charged particles, 
the EFR for the short-range interaction should be modified. Such modification 
generates additional terms in the ERF. These terms depend only on the Coulomb interaction and may far exceed, in the absolute value, the informative part of the ERF containing the phase shifts. This fact hampers the practical procedure of the analytic continuation and affects its accuracy. It was suggested in Ref.~\cite{Sparen} to use for the analytic continuation the quantity $\Delta_l(E)$ [which is defined below (see Section 2)] rather than the ERF. $\Delta_l(E)$ does not contain the pure Coulomb terms. However, the validity of employing $\Delta_l(E)$ was not obvious, which resulted in some discussion since $\Delta_l(E)$, contrary to the ERF, possesses an essential singularity at $E=0$.

In the present work, within the exactly-solvable model, we consider the problem of the analytic continuation from $E>0$ to $E<0$ of the various quantities characterising the elastic scattering in the presence of the Coulomb interaction. The interaction is described by the sum of the square well and  Coulomb potentials.  Various representations of the Coulomb wave functions and corresponding expressions for the partial-wave scattering amplitudes and the ERF are 
considered and the most effective algorithm of analytic continuation is identified. 
It is shown that, although function $\Delta_l(E)$ possesses the essential singularity at $E=0$, it can be nevertheless analytically continued along the real axis of $E$ to the 
region of negative energies. 

The paper is organized as follows. Section 2 contains the general formalism of the elastic scattering for the superposition of the short-range and Coulomb interactions which is necessary for the subsequent discussion. The  various versions of the specific expressions of the scattering phase shifts in the case of the square-well short-range potential are considered in Section 3. The results of calculations within the used model are presented in Section 4.  

We use the system of units in which $\hbar=c=1$ throughout the paper.

\section{General formalism of scattering in the presence of the Coulomb interaction}

The full amplitude of the elastic scattering of particles $b$ and $c$ in the presence of the Coulomb and short-range (nuclear) interactions is written as the 
sum of the pure Coulomb and Coulomb-nuclear amplitudes (the Coulomb interaction is taken to be repulsive):
\begin{align}\label{total}
f(\vec k)&=f_C(\vec k)+f_{NC}(\vec k), \\
\label{fC}
f_C(\vec k)&=\sum_{l=0}^\infty(2l+1)\frac{\exp(2i\sigma_l)-1}{2ik}P_l(\cos\theta), \\
\label{fNC}
f_{NC}(\vec k)&=\sum_{l=0}^\infty(2l+1)\exp(2i\sigma_l)\frac{\exp(2i\delta_l)-1}{2ik}P_l(\cos\theta).
\end{align}
Here $\vec k$ is the relative momentum of $b$ and $c$, $\theta$ is the c.m. scattering angle,   
$\sigma_l=\arg\,\Gamma(l+1+i\eta)$  
 and $\delta_l$ are the pure Coulomb and Coulomb-nuclear phase shifts, respectively, and $\Gamma(z)$ is the Gamma function.
\begin{equation}\label{eta}
\eta =Z_bZ_ce^2\mu/k
\end{equation}
is the Coulomb (Sommerfeld) parameter for the $b+c$ scattering state with the relative momentum $k=\sqrt{2\mu E}$,
$\mu$, $Z_be$, and $Z_ce$ are the reduced mass and the electric charges of $b$ and $c$.

The behavior of the Coulomb-nuclear partial-wave amplitude $f_l=(\exp(2i\delta_l)-1)/2ik$ is irregular near 
$E=0$. Therefore, one has to introduce the renormalized Coulomb-nuclear partial-wave amplitude $\tilde f_l$ 
\cite{Hamilton,BMS,Konig}
\begin{equation}\label{renorm}
\tilde f_l=\exp(2i\sigma_l)\,\frac{\exp(2i\delta_l)-1}{2ik}\,\left[\frac{l!}{\Gamma(l+1+i\eta)}\right]^2e^{\pi\eta}.
\end{equation}
Eq.~(\ref{renorm}) can be rewritten as 
\begin{equation}\label{renorm1}
\tilde f_l=\frac{\exp(2i\delta_l)-1}{2ik}C_l^{-2}(\eta),
\end{equation}
where $C_l(\eta)$ is the Coulomb penetration factor (or Gamow factor)
\begin{align}\label{C}
C_l(\eta)&=\left[\frac{2\pi\eta}{\exp(2\pi\eta)-1}v_l(\eta)\right]^{1/2}, \\ 
v_l(\eta)&=\prod_{n=1}^{l}(1+\eta^2/n^2)\;(l>0),\quad v_0(\eta)=1.
\end{align}
The amplitude $\tilde f_l$ can be expressed in terms of the Coulomb-modified ERF $K_l(E)$ \cite{Hamilton, Konig}
\begin{align} 
\label{fK}
\tilde f_l&=\frac{k^{2l}}{K_l(E)-2\eta k^{2l+1}h(\eta)v_l(\eta)}\\ 
&=\frac{1}{kC_l^2(\eta)(\cot\delta_l-i)} \\ 
&=\frac{1}{v_l^2\Delta_l(E)-ikC_l^2(\eta)},
\end{align}   
where
\begin{align}\label{scatfun}
K_l(E)&= k^{2l+1} \left[ C_l^2(\eta)(\cot\delta_l-i) + 2 \eta h(k)v_l(\eta) \right],\\ 
h(\eta) &= \psi(i\eta) + \frac{1}{2i\eta}-\ln(i\eta), \\ 
\Delta_l(E)&=kC_0^2(\eta)\cot\delta_l, 
\end{align}
and $\psi(x)$ is the digamma function.

It was shown in \cite{Hamilton} that function $K_l(E)$ defined by (\ref{scatfun}) is analytic near $E=0$ and can be expanded into Taylor series in $E$. In the absence of the Coulomb interaction ($\eta=0$) $K_l(E)=k^{2l+1}\cot\delta_l(k)$.

It should be noted that amplitude $\tilde f_l(E)$ possesses the essential singularity at $E=0$. Nevertheless,  
analytic properties of ${\tilde f}_{l}$ on the real axis of the physical sheet of $E$  are analogous to the one of the partial wave scattering amplitude for the short-range potential
and it can be analytically continued into the negative energy region.

If the $b+c$ system involves the bound state $a$ with the binding energy $\varepsilon=\varkappa^2/2\mu>0$, then the amplitude $\tilde f_l$ has a pole at $E=-\varepsilon$. The residue of $\tilde f_l$ at this point is expressed in terms of the ANC 
$C_l$ \cite{BMS}
\begin{align}\label{res2}
{\rm res}\tilde f_l(E)|_{E=-\varepsilon}&=\lim_{\substack{E\to -\varepsilon}}[(E+\varepsilon)\tilde f_l(E)] \\
&=
-\frac{1}{2\mu}\left[\frac{l!}{\Gamma(l+1+\eta_b)}\right]^2C_l^2 ,
\end{align}
where $\eta_b=Z_bZ_ce^2\mu/\varkappa$ is the Coulomb  parameter for the $b+c$ bound state $a$.

\section{Phase shifts for the sum of the square well and Coulomb potentials}

The square well potential is of the form
\begin{equation}\label{potential}
V(r)=\left\{ \begin{matrix}-V_0\ 0\le r\le R\\
0\ r > R \end{matrix} \right. ,
\end{equation}
where $r$ is the distance between interacting particles, $R$ is the radius of the square well, and $V_0>0$ is its depth.

The Schr\"odinger equation describing the system under consideration is 
\begin{equation}\label{cul}
\frac{d^2 u_l(r)}{dr^2} + 2\mu \left[ E- \frac{l(l+1)}{2\mu r^2} -\frac{Z_b Z_c e^2}{r} - V(r) \right] u_l(r) 
= 0. 
\end{equation}
Let us introduce the notations: $\alpha_1=Z_b Z_c e^2 \mu$, $\eta_1=\alpha_1/K$, $K=\sqrt{2\mu(E+V_0)}$. The solution of Eq.~(\ref{cul}) in the inner ($r<R$) and external ($r>R$) regions are given by
\begin{align}\label{R_in}
&R^{in}_l(r)= \frac{u_l(r)}{r} = const \frac{F_l(\eta_1, Kr)}{Kr} , \\
\label{R_out}
&R^{ext}_l(r) = \frac{u_l(r)}{r}, \\ 
&u_l(r)= A_l \left[\chi^{(-)}_l(\eta, kr) - S_l \chi^{(+)}_l(\eta, kr) \right],\\ 
&\chi^{(\pm)}_l(\eta, kr)=G_l(\eta, kr)\pm F_l(\eta, kr), \\ 
&S_l=e^{2i\delta_l}.
\end{align}
In Eqs.~(\ref{R_in}) and (\ref{R_out}) $F_l(\eta, \rho)$ and $G_l(\eta, \rho)$ are the regular and irregular Coulomb functions, respectively \cite{DLMF}. If the Coulomb interaction is turned off ($\eta=0$), then 
\begin{equation}
F_l(0, kr) = kr j_l(kr), \quad G_l(0, kr) = - kr y_l(kr),
\end{equation}
where $j_l(x)=\sqrt{\pi/2x}J_{l+1/2}(x)$ and $y_l(x)=\sqrt{\pi/2x}Y_{l+1/2}(x)$ are spherical Bessel and Neumann functions, respectively.

Now we introduce the functions
\begin{align}\label{hat}
\hat F_{l,\eta}(k,r)  &= F_l(\eta, kr)/kr, \\ 
\hat G_{l,\eta}(k,r) &= -G_l(\eta, kr)/kr, \\
\label{tilde}
\tilde F_{l,\eta}(k,r) &= \hat F_{l,\eta}(k,r) /(k^l C_l(\eta)), \\ 
\tilde G_{l,\eta}(k,r) &= \hat G_{l,\eta}(k,r)k^{l+1} C_l(\eta), \\
\label{Gminus}
\tilde G_{l,\eta}^{(-)}(k,r) &=  k^{l+1} C_l(\eta) \left[ \hat G_{l,\eta}(k,r) - i \hat F_{l,\eta}(k,r) \right] \nonumber \\
&=  \tilde G_{l,\eta}(k,r) - ik^{2l+1}C_l^2(\eta) \tilde F_{l,\eta}(k,r).
\end{align}
In Eqs.~(\ref{tilde})-(\ref{Gminus}) the penetration factor $C_l(\eta)$ is defined by (\ref{C}).

Note that $\tilde F_l(k,r)$ is regular at $E=k^2/2\mu=0$, whereas $\tilde G_l(k,r)$ possesses the Coulomb essential singularity  at 
$E=0$ and behaves irregularly at $E\to-0$ \cite{Humblet}. As to $\tilde G_l^{(-)}$, it is a smooth function on the real axis of $E$. 

The phase shifts $\delta_l$ are found from the condition of equality of logarithmic derivatives of $R_l^{in}(r)$ and $R_l^{ext}(r)$ at 
$r=R$:
\begin{equation}\label{R=r}
\frac{1}{R_l^{in}(R)}\frac{dR_l^{in}(R)}{dR}=\frac{1}{R_l^{ext}(R)}\frac{dR_l^{ext}(R)}{dR}.
\end{equation}
In this equation and hereafter $d\psi(R)/dR\equiv d\psi(r)/dr|_{r=R}$.

Using Eqs.~(\ref{R_in}), (\ref{R_out}), and (\ref{R=r}) we get
\begin{align}
\label{cotdelta}
\cot\delta_l  & \nonumber \\
=&\dfrac{\dfrac{d\hat G_{l,\eta}(k,R)}{dR} \hat F_{l,\eta_1}(K,R)
- \dfrac{d\hat F_{l,\eta_1}(K,R)}{dR} \hat G_{l,\eta}(k,R)} 
{\dfrac{d\hat F_{l,\eta}(k,R)}{dR} \hat F_{l,\eta_1}(K,R)
- \dfrac{d\hat F_{l,\eta_1}(K,R)}{dR} \hat F_{l,\eta}(k,R)} .
\end{align}

\section{Effective range function}

According to Eqs.~(\ref{scatfun}) and (\ref{cotdelta})  the part of the ERF depending on phase shifts is of the form
\begin{align}\label{kind1}
&k^{2l+1} C_l^2(\eta) \cot\delta_l = 
k^{2l+1} C_l^2(\eta) \nonumber \\
&\times \left[
\dfrac {\dfrac{d\hat G_{l,\eta}(k,R)}{dR} \hat F_{l,\eta_1}(K,R)
- \dfrac{d\hat F_{l,\eta_1}(K,R)}{dR} \hat G_{l,\eta}(k,R)}
{\dfrac{d\hat F_{l,\eta}(k,R)}{dR} \hat F_{l,\eta_1}(K,R)
- \dfrac{d\hat F_{l,\eta_1}(K,R)}{dR} \hat F_{l,\eta}(k,R)} \right] .
\end{align}
We transform (\ref{kind1}) using the modified Coulomb functions $\tilde F_{l,\eta}
$, 
and $\tilde G_{l,\eta}
$ 
introduced earlier to get
\begin{align}\label{kind2}
k^{2l+1} & C_l^2(\eta) \cot\delta_l   \nonumber \\
=&\dfrac {\dfrac{d\tilde G_{l,\eta}(k,R)}{dR} \tilde F_{l,\eta_1}(K,R)
- \dfrac{d\tilde F_{l,\eta_1}(K,R)}{dR} \tilde G_{l,\eta}(k,R)}
{\dfrac{d\tilde F_{l,\eta}(k,R)}{dR} \tilde F_{l,\eta_1}(K,R)
- \dfrac{d\tilde F_{l,\eta_1}(K,R)}{dR} \tilde F_{l,\eta}(k,R)}  .
\end{align}
The denominator in (\ref{kind2}) does not possess the Coulomb singularities at $E> -V_0$, however, the function $\tilde G_l(k, R)$ is singular at $E=0$ (see Section 3). The singularities of  $k^{2l+1} C_l^2(\eta) \cot\delta_l$ can be singled out using  
Eq.~(\ref{Gminus}) as
\begin{align}\label{kind3}
k^{2l+1} &C_l^2(\eta) \cot\delta_l  \nonumber \\
=&\dfrac {\dfrac{d\tilde G_{l,\eta}^{(-)}(k,R)}{dR} \tilde F_{l,\eta_1}(K,R)
- \dfrac{d\tilde F_{l,\eta_1}(K,R)}{dR} \tilde G_{l,\eta}^{(-)}(k,R)}{\dfrac{d\tilde F_{l,\eta}(k,R)}{dR} \tilde F_{l,\eta_1}(K,R)
- \dfrac{d\tilde F_{l,\eta_1}(K,R)}{dR} \tilde F_{l,\eta}(k,R)} \nonumber \\
& + i k^{2l+1} C^2_l(\eta). 
\end{align}
The first term in the right-hand side of Eq.~(\ref{kind3}) is regular at $E=0$ whereas the second term has an essential singularity at this point.

In what follows the specific properties of different parts of the ERF and the partial-wave scattering amplitude will be illustrated by numerical calculations applying to the $d+\alpha$ system in the $S$ state ($l=0$). This system involves one bound state corresponding to the ground state of the $^6$Li nucleus. An accurate bound-state information is required, e.g. for modeling astrophysical $\alpha+d$ $\rightarrow$ $^6$Li + $\gamma $ radiative capture reaction \cite{IY00,MBI11,MSB16,TKTB16}. The following parameters are used in the calculations: $m_\alpha=3755.58$ MeV, $m_d=877.79$ MeV,  
$V_0=7.64386$ MeV, $R=3.73473$ fm. The values of $V_0$ and $R$ were fitted to the values of the binding energy $\varepsilon=2.409$ MeV and the dimensionless ANC $\tilde C_0=2.29$ of $^6$Li in the $d+\alpha$ channel obtained by solving the Faddeev equations for $^6$Li without  the Coulomb interaction \cite{BKSSK}. 

\begin{figure}[htb]
\center{\includegraphics[width=\columnwidth]{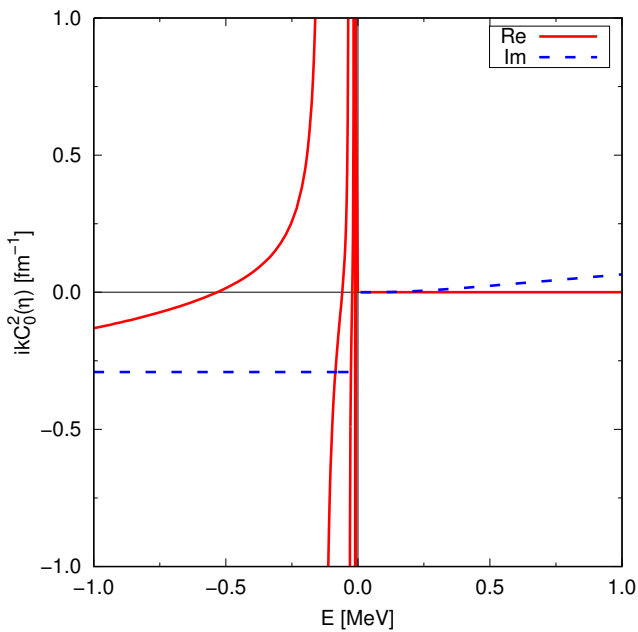}}
\caption{Functional dependence of $i k C_0^2(\eta)$ on energy $E$.}
\label{pic1}
\end{figure}

\begin{figure}[htb]
\center{\includegraphics[width=\columnwidth]{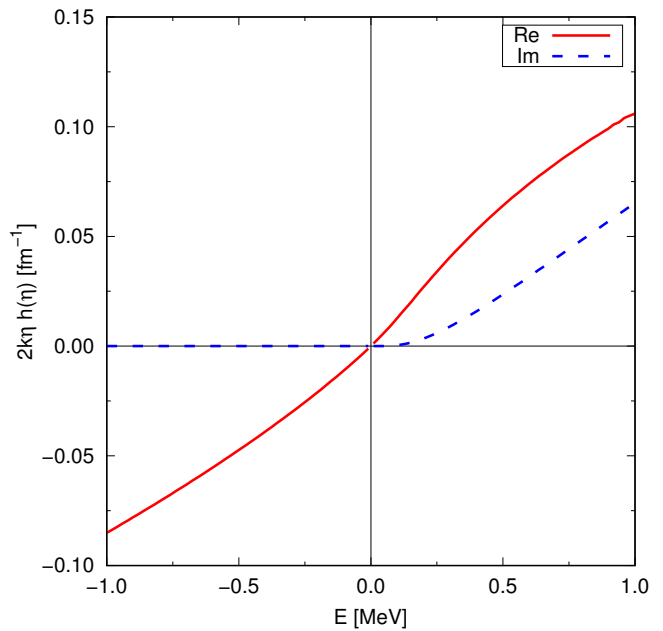}}
\caption{Functional dependence of $2k \eta h(\eta)$ on energy $E$.} 
\label{pic2}
\end{figure}

Consider first the features of the function $i k^{2l+1}  C_l^2(\eta)$ which enters the expression for the ERF.
The energy dependence of this function at $l=0$ is presented in Fig.~\ref{pic1}. It is seen that the imaginary part of $i k^{2l+1}  C_l^2(\eta)$ is constant at $E<0$. This property can be rigorously proved.  Making use of the definition (\ref{C}) for $C_l(\eta)$ and the notations 
$\eta_b=\alpha_1/\varkappa$, $\alpha_1=Z_bZ_ce^2/\mu$, one obtains at $E<0$
\begin{align}\label{iC0}
i k C_0^2(\eta) &= \frac{2\pi i \alpha_1}{e^{-2\pi i \eta_b} -1} = \frac{2\pi i \alpha_1 e^{\pi i \eta_b}}{e^{-\pi i \eta_b} -
e^{\pi i \eta_b}} \nonumber \\
&= -\pi \alpha_1 \frac{e^{\pi i \eta_b}}{\sin \pi \eta_b} =-\pi \alpha_1 \cot \pi \eta_b -\pi \alpha_1 i. 
\end{align}
It follows from (\ref{iC0}) that Re$[i k C_0^2(\eta)]$ possesses poles at $\eta_b = 1,2, \cdots $ corresponding to $\varkappa=\alpha_1/n$ ($n=1,2, \cdots$),  and zeroes at $\eta_b = 1/2, 3/2, 5/2 \cdots $ corresponding to  $\varkappa=2 \alpha_1/n$  
($n=1,3,5, \cdots$).
Another function that appears in the expression for the ERF and helps regularize it, is $2k \eta h(\eta)$. Its plot is shown 
in Fig.~\ref{pic2}.

Now we demonstrate
that all three expressions given by Eqs. (\ref{kind1}), (\ref{kind2}) and (\ref{kind3}) can be used to calculate  
 $k^{2l+1} C_l^2(\eta) \cot\delta_l$ and then compare the accuracy achieved by these expressions. To this end, the ERF $K_0(E)$ of Eq.~(\ref{scatfun}) was calculated at $E=5$ MeV and $E=-5$ MeV. The calculations were performed to ten significant digits. The results of the calculations are presented in Table~1. 
The exact ERF should be real both at positive and negative energies. It is seen from the table that all three versions lead to similar results. However, as to the computational side, the versions (\ref{kind2}) and (\ref{kind3}) exploiting the modified Coulomb functions are more preferable since they lead to more accurate results at negative energies. The advantage of these versions increases further as $E$ approaches zero. Note that the version (\ref{kind3}) is the most accurate out of all three versions. All numerical results for $\cot\delta_l$ discussed below are obtained using this version.

\begin{table*}[htb]
\centering
\label{Tab1} 
\caption{Effective-range functions calculated using Eqs.~(\ref{kind1}), (\ref{kind2}) and (\ref{kind3}).}
\begin{tabular}{ccc} 
\hline \hline
version & $K_0(5)$ & $K_0(-5)$\\ 
\hline 
Eq.~(\ref{kind1}) & $0.3120231954-i 0.14 \times 10^{-8}$ & $-0.2802558905-i 0.63 \times 10^{-4}$ \\ 
Eq.~(\ref{kind2}) & $0.3120231949-i 0.40 \times 10^{-9}$ & $-0.2802915315- i 0.10 \times 10^{-9}$ \\ 
Eq.~(\ref{kind3}) & $0.3120231949- i 0.40 \times 10^{-9}$ & $-0.2802915316+i 0.14 \times 10^{-10}$ \\ 
\hline \hline
\end{tabular} 
\end{table*}

\begin{figure}[htb]
\center{\includegraphics[width=\columnwidth]{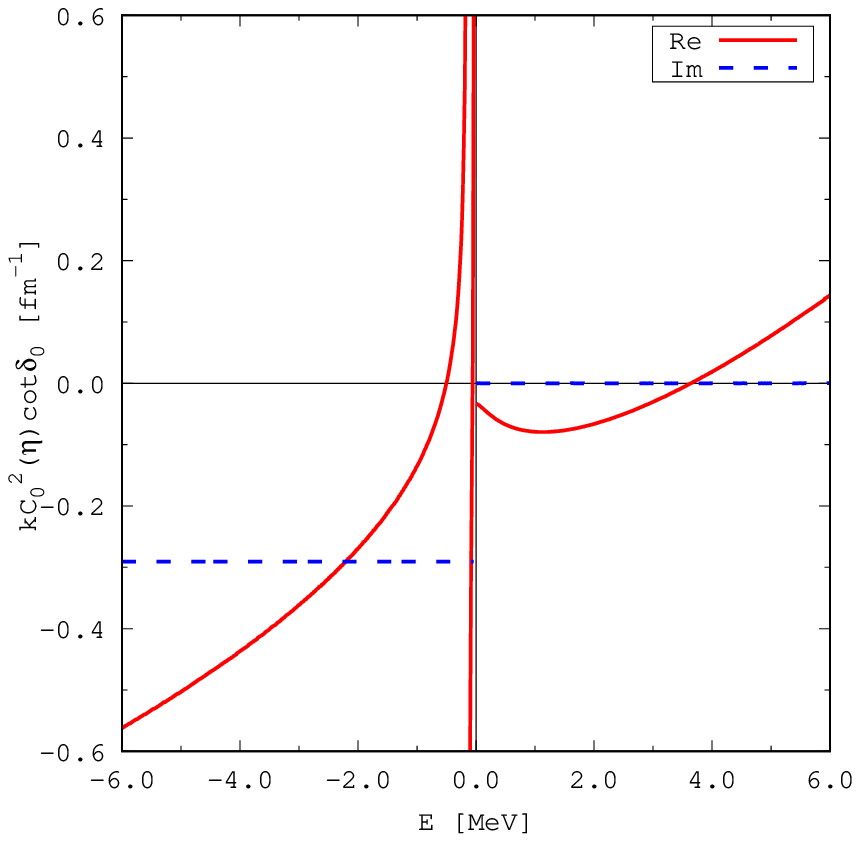}}
\caption{Functional dependence of $k C_0^2(\eta) \cot\delta_0$ on energy $E$.}  
\label{pic3}
\end{figure}

The most important part of the ERF is the function $k^{2l}v_l^2\Delta_l(E) = k^{2l+1} C_l^2(\eta) \cot\delta_l$.
 Its energy dependence at $l=0$ is presented in Fig.~\ref{pic3}. The function is complex at $E<0$ and displays the irregularities which are concentrated near zero from the left. It is seen that 
Im[$k C_0^2(\eta) \cot\delta_0]$ is constant at $E<0$. Moreover, this constant is exactly equal to Im[$ik C_0^2(\eta)$] at $E<0$ [see 
Fig.~\ref{pic1} and Eq.~(\ref{iC0})].

\begin{figure}[htb]
\center{\includegraphics[width=\columnwidth]{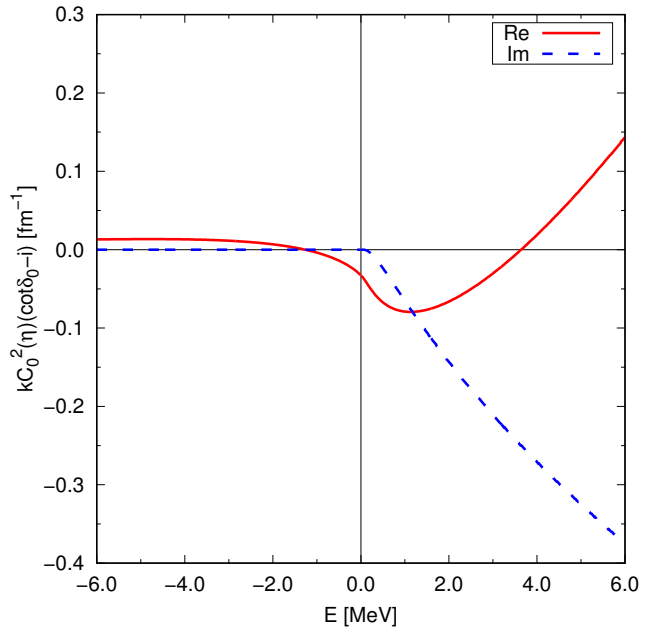}}
\caption{Functional dependence of $kC_0^2(\eta) (\cot\delta_0 -i)$ on energy $E$.}  
\label{pic4}
\end{figure}

Consider now  the function $D_l(E)\equiv k^{2l+1} C_l^2(\eta) (\cot\delta_l - i)$. Its plot is shown in Fig.~\ref{pic4} for $l=0$. The function $D_0(E)$ 
is complex at positive energies and real at negative energies.
Furthermore, as is seen from Figs.~\ref{pic1}, \ref{pic3} and \ref{pic4}, the irregularities of functions $k C_0^2(\eta) \cot\delta_0$ and 
$i kC_0^2$ at $E\to-0$ disappear in the function $D_0(E)$, which is their difference.  
Note that $D_0(E)$ is the inverse of the renormalized partial-wave Coulomb-nuclear amplitude $\tilde f_0$, see Eq.~ (\ref{fK}): $D_0(E)=(\tilde f_0)^{-1}$ [generally $D_l(E)=k^{2l}(\tilde f_l)^{-1}$].

 Thus the behavior of $D_0(E)$ shown in Fig.~\ref{pic4} corroborates the above assertion that on the real $E$ axis 
 the analytical properties of
 the amplitude 
 $\tilde f_0$ 
 are similar to those of
 of the  partial-wave amplitude of scattering from a short-range potential. Hence ${\tilde f}_{0}$ can be analytically continued to the negative energy region. However, one should not forget that both $D_0(E)$ and $\tilde f_0(E)$ possess the essential singularity at $E=0$.

\begin{figure}[htb]
\center{\includegraphics[width=\columnwidth]{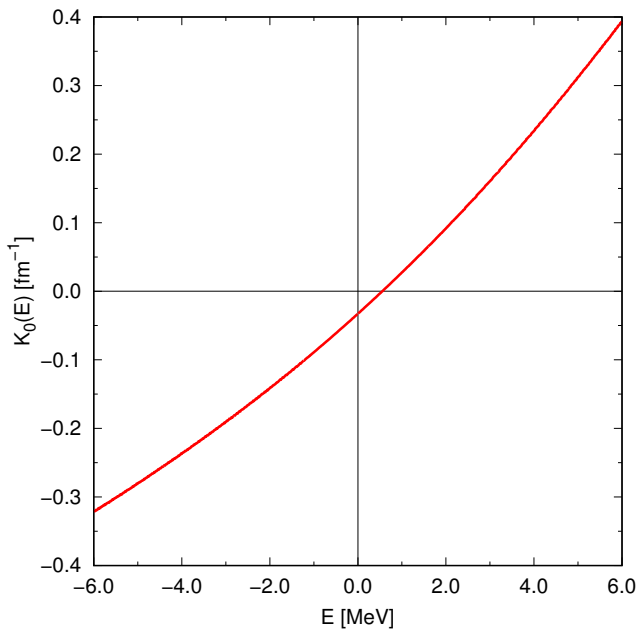}}
\caption{Dependence of ERF $K_0(E)$ on energy $E$. The imaginary part is 0.} 
\label{pic5}
\end{figure}

Finally, we consider the full ERF $K_l(E)$. The energy dependence of $K_0(E)$ is displayed in Fig.~\ref{pic5}. The ERF is real for all real values of $E$ and is not singular at $E=0$.

The 
pole of the amplitude $\tilde f_0$ was found using the requirement 
$D_0(E)\equiv k C_0^2(\eta) (\cot\delta_0 - i)=0$. The zero of $D_0(E)$ corresponds to the point where the red line crosses the 
negative x-axis, see Fig.\ref  {pic4}. For the $d+\alpha$ system we obtain the following values of the binding energy $\varepsilon$ and the dimensionless ANC $\tilde C_0$: $\varepsilon$=1.268 MeV, $\tilde C_0= C_0/\sqrt{2\varkappa} =2.683$. These values coincide with the values of 
$\varepsilon$ and 
$\tilde C_0$ obtained from direct solution of the Schr\"odinger equation (\ref{cul}) for the bound state of the $d+\alpha$ system. This agreement confirms the validity of the employed procedure of the analytic continuation of the scattering characteristics from positive to negative energies.

Let us remind that if the Coulomb interaction is turned off, then $\varepsilon$=2.409 MeV, $\tilde C_0= C_0/\sqrt{2\varkappa} =2.29$. The experimental value of $\varepsilon$ is 1.47 MeV. The analytic approximation of the experimental $d+^4$He scattering phase shifts with subsequent continuation to negative energies results in $\tilde C_0$ =2.93 \cite{BKSSK}. 

Recently in a number of works experimental data on the elastic scattering have been used to get the information about bound states. 
The following procedure is usually used (see, e.g. \cite{SpCaBa,IrOr}). The values of the ERF $K_l(E)$ at $E>0$ obtained by the phase-shift analysis of experimental data are approximated by some analytic function, say, by several first terms of the effective-range expansion or by the Pad\'e approximant. The approximating function obtained in such a way, as well as the exact function $K_l(E)$ do not possess singularities at $E=0$. Hence it could be continued analytically to the negative energy region. After that, using (\ref{fK}) one can find the location of the pole of the amplitude $\tilde f_l$ corresponding to the bound state and the residue of $\tilde f_l$ at that pole in terms of which the value of the ANC is expressed.

\begin{figure}[htb]
\center{\includegraphics[width=\columnwidth]{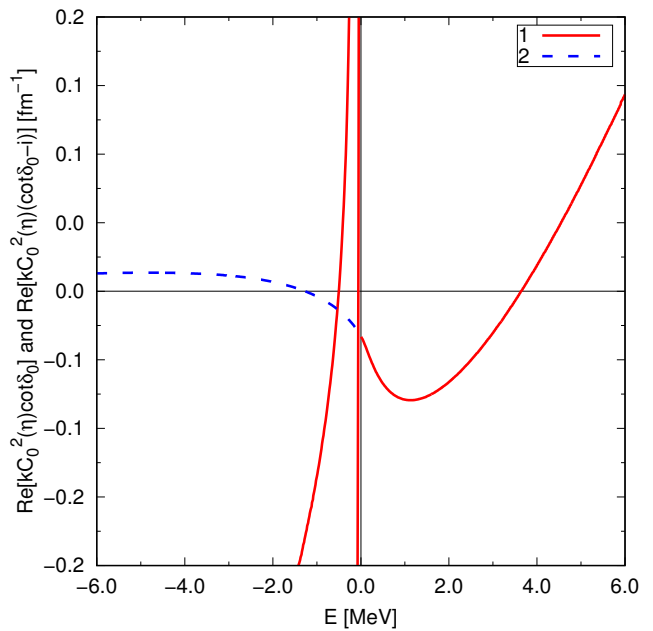}}
\caption{Functional dependence of ${\rm Re}[k C_0^2(\eta) \cot\delta_0]$ (line 1) and ${\rm Re}[k C_0^2(\eta) (\cot\delta_0 -i)]$ (line 2) on energy $E$. For $E>0$ they coincide.} 
\label{pic6}
\end{figure}

A somewhat different procedure was used in Ref.~\cite{Sparen} 
in order to obtain the information about the bound states of the $^{16}$O nucleus from the data on the elastic $\alpha-^{12}$C scattering:
 instead of $K_l(E)$,  function $\Delta_l(E)=k^{2l+1}C_l^2(\eta)\cot\delta_l$
 is approximated and continued to the region $E<0$. Our results presented in Fig.~\ref{pic3} show that $\Delta_l(E)$ is a smooth function of $E$ at $E>0$, however, its behaviour is irregular near $E=0$ at $E<0$. On the other hand, at $E>0$ function $\Delta_l(E)$ 
coincides with the quantity $\tilde\Delta_l(E)\equiv {\rm Re}[k^{2l+1} C_l^2(\eta) (\cot\delta_l-i)]$. As is seen from Figs.~\ref{pic4} and 
\ref{pic6}, $\tilde\Delta_l(E)$ is a smooth function of $E$ both at $E>0$ and at $E<0$. Therefore $\tilde\Delta_l(E)$ can be considered as the analytic continuation of the function $\Delta_l(E)$ defined at $E>0$ to the negative-energy region. This circumstance may serve as the justification of the procedure suggested in \cite{Sparen}.

\section{Conclusions}

In the present paper the explicit analytic expressions have been derived for the ERF 
and the partial-wave scattering amplitude in the case of the interaction given by the sum of a short-range square-well and the Coulomb potentials. These expressions have been transformed into the forms convenient for the analytic continuation to the negative-energy region. The analytic properties of separate terms entering the expressions for the ERF and the scattering amplitude have been investigated.  

It is demonstrated that  function $\Delta_l(E)$ suggested in Ref.~\cite{Sparen} can be used to obtain information about bound state properties. In spite of having the essential singularity at $E=0$ function $\Delta_l(E)$ can be analytically continued from the positive to the negative energy region along the real $E$ axis. For instance, function $f(z) = \exp(-1/z^2)$ provides an example of such functions. Indeed, $f(z)$ possesses the essential singularity at $E=0$. Nevertheless, the function itself and all its derivatives are smooth functions on the real axis including $z=0$. Using $\Delta_l(E)$ rather than the ERF $K_l(E)$ might be preferable since $K_l(E)$, in contrast to $\Delta_l(E)$, contains a pure Coulomb term which may far exceed, in the absolute value, the term containing the information about the phase shifts.

Note that all qualitative results obtained in the present work do not depend on the specific values of the parameters of the potential used in the numerical calculations. Moreover, though all calculations were performed for $l=0$, the inferences made should be valid for arbitrary $l$.

It is interesting to note that if the Coulomb interaction is turned off ($\eta=0$), then $\tilde\Delta_l(E)$ evidently loose the Coulomb essential singularity but  acquires a square-root type singularity  at $E=0$ which is absent at $\eta\ne 0$. This singularity corresponds to the normal threshold of the scattering amplitude. It follows from our calculations that, if one substitutes 
$\gamma Z_bZ_c$  for $Z_bZ_c$ in Eq.~(\ref{eta}), then at $\gamma\to 0$ the first derivative of $\tilde\Delta_0(E)$ tends to $-\infty$.  
One may conclude that the behavior of $\tilde\Delta_l(E)$ in the vicinity of $E=0$ is more smoother at larger values of $\eta$,  that is, at larger values of $Z_bZ_c\mu$. It means that the procedure suggested in \cite{Sparen} is more effective for heavier nuclei. 

Finally, the results presented in this paper should be useful for researchers working in the effective field theory \cite{PRS00,HHK08,KLH13} as well
since they fit the elastic scattering data at positive energies, and here we investigate a possibility of extrapolating the data to the bound-state pole.

\section*{Acknowledgements}
This work was supported by the Russian Science Foundation (Grant No. 16-12-10048). 
A.S.K. acknowledges a support from the Australian Research Council. 
A.M.M. acknowledges that this material is based upon work supported by the US Department of Energy, Office of Science, Office of Nuclear Science, under Grant No. DEFG02- 93ER40773. It is also supported  by the US National Science Foundation under Grant No. PHY-1415656.


\end{document}